\documentstyle[12pt,rotate]{article}
\input epsf.sty

\newcommand{\inclfig}[2]{\mbox{\epsfxsize=#1cm \epsfbox{#2.ps}}}

\newcommand{\Bx}{x_{\rm Bj}}
\newcommand{\cF}{{\cal F}}
\newcommand{\cG}{{\cal G}}
\newcommand{\cT}{{\cal T}}
\newcommand{\cO}{{\cal O}}
\newcommand{\cQ}{{\cal Q}}
\newcommand{\MS}{{$\overline{\mbox{MS}}$}}

\begin{document}

\begin{titlepage}

\centerline{\large \bf Deeply virtual Compton scattering in
                       next-to-leading order.}

\vspace{15mm}

\centerline{\bf A.V. Belitsky\footnote{Alexander von Humboldt Fellow.},
             D. M\"uller, L. Niedermeier, A. Sch\"afer}

\vspace{15mm}

\centerline{\it Institut f\"ur Theoretische Physik, Universit\"at
               Regensburg}
\centerline{\it D-93040 Regensburg, Germany}

\vspace{20mm}

\centerline{\bf Abstract}

\hspace{0.5cm}

We study the amplitude of deeply virtual Compton scattering in
next-to-leading order of perturbation theory including the two-loop
evolution effects for different sets of skewed parton distributions
(SPDs). It turns out that in the minimal subtraction scheme the
relative radiative corrections are of order 20-50\%. We analyze the
dependence of our predictions on the choice of SPD, that will allow
to discriminate between possible models of SPDs from future high
precision experimental data, and discuss shortly theoretical
uncertainties induced by the radiative corrections.

\vspace{6cm}


\vspace{0.5cm}


\end{titlepage}

Virtual Compton scattering (VCS) in a light cone dominated region,
--- usually referred to as the deeply virtual Compton scattering (DVCS)
\cite{MueRobGeyDitHor94}-\cite{Rad96}, --- is a favourable process to
get access to the so-called skewed parton distributions (SPDs)
\cite{MueRobGeyDitHor94}-\cite{Rad97}. The latter can be viewed as a
generalization of the conventional parton densities as measured in deep
inelastic scattering (DIS). They contain complementary information
about the internal structure of hadrons, e.g.\ the total angular
momentum carried by quarks and gluons \cite{Ji97}.

The study of SPDs promises to become one of the main issues of hadronic
physics for the next decade. This marks general trends in the development
of experimental techniques towards exclusive reactions at the facilities
like CERN and DESY. In general, it will be a delicate task to measure DVCS,
since it is contaminated by the Bethe-Heitler (BH) process. However, it is
rather encouraging that a first DVCS signal has already been detected by
the ZEUS collaboration \cite{Sau99}. In addition, azimuthal, spin and
charge asymmetries \cite{FraFreStr97} allow to get access to the
interference term between DVCS and BH processes and this should allow to
separate real and imaginary parts of the DVCS amplitude. Such measurements
would pin down plausible shapes of SPDs.

The theoretical goal is, therefore, to push our understanding of SPDs
to the level reached nowadays for the usual parton distribution
functions. One of the most important questions is the $Q^2$-evolution
of SPDs. In earlier papers we gave a complete solution at the NLO level
\cite{BelMueNieSch98}. In the present paper we study the r\^ole of the
NLO radiative corrections for the DVCS amplitude quantitatively in the
region $\Bx > 10^{-2}$ including the effects of scaling violations. Our
aim is to decide whether an extension to NNLO is necessary to obtain a
stable interpretation of a given set of DVCS data. For a first rough
comparison with experimental measurements NLO is probably fine but to
judge the potential of SPDs measurement the precise size of the
corrections as well as the uncertainties involved have to be known.

The VCS amplitude is defined in terms of the time ordered product of two
electromagnetic currents
\begin{equation}
T_{\mu\nu} (P, q_1, q_2)
= i \int dx e^{ixq}
\langle P_2, S_2 |
T \left\{ j_\mu ( x/2 ) j_\nu( - x/2 ) \right\}
| P_1, S_1 \rangle .
\end{equation}
Here $P_1$ and $P_2$ are the momenta of the initial and the final hadrons,
respectively. The incoming photon with momentum $q_1$ has a large
virtuality. The scaling variables, $\xi$ and $\eta$, which allow to
describe different ``two-photon'' processes in the light-cone dominated
region are introduced as follows \cite{MueRobGeyDitHor94}
\begin{eqnarray}
\label{Variables}
&&\xi \equiv \frac{Q^2}{P q},
\quad
\eta \equiv \frac{\Delta q}{P q},
\quad\mbox{where} \\
&&Q^2 = -q^2 = - \frac{1}{4} ( q_1 + q_2 )^2,
\quad
P = P_1 + P_2,
\quad
\Delta = P_1 - P_2 = q_2 - q_1 . \nonumber
\end{eqnarray}
In DVCS kinematics $- q_1^2 \gg m_{\rm hadron}^2$, $q_2^2 = 0$ and we
have in fact only one scaling variable $\xi$ since $\eta = \xi \left(1
- \frac{\Delta^2}{4 Q^2} \right) \approx \xi$. From the experimental
point of view it is more appropriate to work with the variables $- q_1^2$
and $\Bx \equiv - q_1^2/2(P_1 q_1)$, which are related to the variables
(\ref{Variables}) by
\begin{eqnarray}
Q^2 = - \frac{1}{2} q_1^2 \left( 1 - \frac{\Delta^2}{2q_1^2} \right)
\approx - \frac{1}{2} q_1^2 ,
\quad
\xi = \frac{ \Bx \left( 1 - \frac{\Delta^2}{2 q_1^2} \right) }
{ 2 - \Bx \left( 1 + \frac{\Delta^2}{q_1^2} \right) }
\approx \frac{ \Bx }{ 2 - \Bx } .
\end{eqnarray}

In the kinematical region we are interested in there are two leading
twist contributions
\begin{eqnarray}
\label{decom-T}
T_{\mu\nu} (P, q_1, q_2)
= - \tilde g^T_{\mu\nu}
\cF_1 ( \xi, \eta = \xi, Q^2, \Delta^2 )
+ i \tilde \epsilon_{\mu\nu q P} \frac{1}{Pq}
\cG_1 ( \xi, \eta = \xi, Q^2, \Delta^2 )
+ \cdots,
\end{eqnarray}
where the ellipsis stand for a leading twist-two contribution coming from
longitudinally polarized photons which, however, does not appear in the
DVCS cross section, as well as higher twist functions. The transverse part
of the metric tensor, denoted by ${g}^T_{\mu\nu}$ and the $\epsilon$-tensor
are contracted with the projection operators ${\cal P}_{\alpha\beta} =
g_{\alpha\beta} - q_{2\alpha} q_{1\beta} / (q_1 q_2)$. Therefore, current
conservation is manifest. Our definitions are chosen in such a way that
in the forward case, i.e.\ $\Delta = 0$, the usual structure functions
measured in DIS are
\begin{equation}
F_1 ( \Bx, Q^2) = \frac{1}{2\pi} {\rm Im} \cF_1 (\xi = \Bx, 0, Q^2, 0),
\quad
g_1( \Bx, Q^2)  = \frac{1}{2\pi} {\rm Im} \cG_1 (\xi = \Bx, 0, Q^2, 0).
\end{equation}

Since the virtuality of the incoming photon is deep in the Euclidean
domain the hadron is probed almost on the light cone $x^2 \approx 0$.
Hence, the amplitude can be straightforwardly treated by a non-local
version of the light-cone operator product expansion \cite{AniZav78}.
More recently, it has been proven that assuming a smooth SPD the
collinear singularities are indeed factorizable at leading twist to
all orders of perturbation theory \cite{Rad97,FreCol98,JiOsb98}.
Therefore, the amplitudes $\cF_1$ and $\cG_1$ factorize in a perturbative
hard scattering amplitude and a SPD $q (t, \eta, \Delta^2, \mu^2)$ as
\begin{equation}
\cT^a = \sum_{Q}
e^2_Q \int_{-1}^1 \frac{d t}{|\eta|}
\Bigg[
{^Q T^a} (\xi, \eta, t, Q^2, \mu^2) \,
{^Q\! q^a} (t, \eta, \Delta^2, \mu^2)
+ \frac{1}{N_f} \frac{1}{\eta}
{^G T^a} (\xi, \eta, t, Q^2, \mu^2) \,
{^G\! q^a} (t, \eta, \Delta^2, \mu^2)\Bigg] ,
\end{equation}
where $\cT^V = \cF_1$, $\cT^A = \cG_1$ and the sum runs over quark
species $Q = u, d, s$ with electrical charge $e_Q$.

The non-perturbative input is concentrated in the SPDs which read in
the Leipzig conventions \cite{MueRobGeyDitHor94}
\begin{eqnarray}
\left\{ { {^Q\!q^V} \atop {^Q\!q^A} } \right\} ( t, \eta )
&=& \int \frac{d\kappa}{2\pi} e^{i \kappa t P_+}
\langle P_2 S_2 |
\bar \psi (- \kappa n)
\left\{ { \gamma_+ \atop \gamma_+ \gamma_5 } \right\}
\psi (\kappa n)
| P_1 S_1 \rangle ,  \\
\left\{ { {^G\!q^V} \atop {^G\!q^A} } \right\} ( t, \eta )
&=& \frac{4}{P_+} \int \frac{d\kappa}{2\pi} e^{i \kappa t P_+}
\langle P_2 S_2 |
G^a_{+ \mu} (-\kappa n)
\left\{ { g_{\mu\nu} \atop i \epsilon_{\mu\nu-+} } \right\}
G^a_{\nu+} (\kappa n)
| P_1 S_1 \rangle .
\end{eqnarray}
For the quark distributions we have omitted for brevity the flavour
indices. A form factor decomposition would give us the functions $H$,
$E$ and $\widetilde H$, $\widetilde E$ for the parity even and odd
sectors, respectively, as introduced in Ref.\ \cite{Ji97}.

Using the spatial parity and scaling properties evident from the
convolution-type formulae of Ref.\ \cite{BelMue97a} we write the
perturbative expansion of $T$ as
\begin{equation}
\xi \ {^i T^a}
= {^i T^{a(0)}}
\left( \frac{\xi}{\eta}, \frac{t}{\eta} \right)
+ \frac{\alpha_s (\mu^2)}{2\pi}
{^i T^{a(1)}}
\left( \frac{\xi}{\eta}, \frac{t}{\eta}, \frac{Q^2}{\mu^2} \right)
+ \cO \left(\alpha_s^2\right) \mp \left( t \to -t \right) ,
\end{equation}
where the ``$-$''(``$+$'') sign corresponds to the $V$($A$) channel.
The tree level coefficient functions are
\begin{equation}
{^Q T^{(0)}} \left( \frac{\xi}{\eta}, \frac{t}{\eta} \right)
= \frac{1}{1 - t/\xi - i \epsilon},
\qquad
{^G T^{(0)}} = 0.
\end{equation}

The hard scattering amplitude ${T^{(1)}}$ can be calculated by making use
of standard methods of perturbative QCD \cite{JiOsb97}. At the same time
there exists an interesting possibility to predict \cite{BelMue97a}
these quantities with the help of the conformal operator product expansion
(COPE). It has been introduced more than two decades ago in Ref.\
\cite{FerGriGat71} and consequently applied to exclusive processes at
leading order \cite{BroFriLepSac80}. The main advantage of the COPE is
that under the assumption of conformal covariance it predicts the Wilson
coefficients of local conformal operators with a given conformal spin
entering the expansion of the product of two currents up to a normalization
constant. The latter is fixed by the known Wilson coefficients of forward
DIS. Since the COPE is valid only for $|\xi| > 1$ we perform a summation
over all conformal partial waves and obtain a representation of the hard
scattering amplitudes as a convolution of kernels. E.g.\ for the
non-singlet case we have
\begin{eqnarray}
\label{nonloc-T}
T (\xi, \eta, t, Q^2, \mu^2)
= \frac{1}{\xi}
F \left( \frac{\xi}{\eta}, \frac{r}{\eta} \right)
\otimes
\left( \frac{Q^2}{\mu^2} \right)^{
V \left( \frac{r}{\eta}, \frac{s}{\eta} \right) }
\otimes
C \left( \frac{s}{\eta}, \frac{t}{\eta}; \alpha_s \right),
\end{eqnarray}
where the convolution is defined for any test functions $\tau_1 (x, y)$
and $\tau_2 (x, y)$ with $\tau_1 \otimes \tau_2 \equiv \int
\frac{d y}{|\eta|} \tau_1 (x, y) \tau_2 (y, z)$. The evolution kernel
$V \left( \frac{r}{\eta}, \frac{s}{\eta} \right)$ and the coefficient
function $C \left( \frac{s}{\eta}, \frac{t}{\eta}; \alpha_s \right)$ are
diagonal w.r.t.\ the conformal waves --- the Gegenbauer polynomials:
\begin{equation}
\label{coef-func}
\eta^{j} C_{j}^{3/2} \left( \frac{t}{\eta} \right)
\otimes
\left\{ { V \atop C }\right\}
\left( \frac{t}{\eta}, \frac{t'}{\eta}; \alpha_s \right)
=
\left\{ { - \frac{1}{2} \gamma_j (\alpha_s)
\atop
\phantom{- \frac{1}{2}} c_j (\alpha_s) }\right\}
\eta^j C_{j}^{3/2} \left( \frac{t'}{\eta} \right) ,
\end{equation}
and their eigenvalues $\gamma_j(\alpha_s)$ and $c_j(\alpha_s)$ coincide
with anomalous dimensions and Wilson coefficients, respectively, which
appear in DIS. A closed form of the function $F$ can be deduced order
by order in $\alpha_s$ from the sum
\begin{equation}
\label{def-F}
F (x, y) = \sum_{j = 0}^\infty
\left( \frac{2 x}{ 1 + x } \right)^{j + 1}
\frac{B(j + 1, j + 2)}{(1 + x)^{\gamma_j/2}}
{_2F_1} \left( \left.
{ 1 + j + {\gamma_j/2}, \ 2 + j + {\gamma_j/2} \atop 4 + 2j + {\gamma_j} }
\right| \frac{2 x}{1 + x} \right) C_{j}^{3/2} (y) .
\end{equation}

The expressions (\ref{nonloc-T}-\ref{def-F}) which are exact in QCD up
to NLO\footnote{Beyond this order the trace anomaly appears and provides
conformal noncovariant terms proportional to $\beta/g [\alpha_s/(2\pi)
+ O(\alpha_s^2)]$.} define the so-called conformal subtraction (CS)
scheme. To obtain the \MS\ results one has to perform a scheme
transformation \cite{BelMue97a,Mue97a}, which is governed by a conformal
anomaly appearing in special conformal Ward identities \cite{BelMue98c}.
Finally, the NLO coefficient functions read
\begin{eqnarray}
{^{Q}\!T}^{(0)}
\!\!\!&=&\!\!\!
\frac{1}{1 - t},
\qquad
{^{Q}\!T^{V(1)}} = {^{Q}\!T^{A(1)}}
- \frac{C_F}{1 + t} \ln \frac{1 - t}{2}, \\
{^{Q}\!T^{A(1)}}
\!\!\!&=&\!\!\!
\frac{C_F}{2 (1 - t)}
\left[ \left( 2 \ln \frac{1 - t}{2} + 3 \right)
\left(
\ln\frac{- q_1^2}{\mu^2} + \frac{1}{2} \ln\frac{1 - t}{2} - \frac{3}{4}
\right)
- \frac{27}{4} - \frac{1 - t}{1 + t} \ln \frac{1 - t}{2}
\right], \\
{^{G}\!T^{V(1)}}
\!\!\!&=&\!\!\!
- {^{G}\!T^{A(1)}}
+ \frac{N_f}{2}
\left[ \frac{1}{1 - t}
\left( \ln\frac{- q_1^2}{\mu^2} + \ln\frac{1 - t}{2} - 2 \right)
+ \frac{\ln\frac{1 - t}{2}}{1 + t} \right] , \\
{^{G}\!T^{A(1)}}
\!\!\!&=&\!\!\!
\frac{N_f}{2} \left[
\left( \frac{1}{1 - t^2} + \frac{\ln\frac{1 - t}{2}}{(1 + t)^2} \right)
\left( \ln\frac{- q_1^2}{\mu^2} + \ln\frac{1 - t}{2} - 2 \right)
- \frac{\ln^2 \frac{1 - t}{2}}{2(1 + t)^2} \right] .
\end{eqnarray}
We have used here the scaling property mentioned above and the fact that
the hard scattering amplitude for DVCS, i.e.\ $\xi = \eta$, is a simple
analytical continuation of the ones for the production of a scalar and
pseudo-scalar mesons, respectively, in the collision of a real and a
highly virtual photon.

To the same accuracy we have to include the two-loop evolution of the
SPDs. Based on the knowledge of NLO anomalous dimensions constructed
from conformal constraints \cite{BelMue98c} we will use the methods
of Ref.\ \cite{BelMueNieSch98} where we have already dealt with scaling
violations in NLO.

For numerical estimates of the DVCS amplitude we expand the hard
scattering amplitude in terms of Legendre polynomials and reexpress
the expansion coefficients in terms of conformal moments. This allows
us to include easily the NLO evolution of the SPDs as described in Ref.\
\cite{BelMueNieSch98}. Unfortunately, this method is reliable only for
moderate values of $\Bx$ since the convergence of the series of
polynomials is not under control for very low $\Bx$. Already for $\Bx
\sim 0.05$ one needs about 180 polynomials.

To model the SPD one can first explore the simplest possibility of
equating it to the usual forward parton density. We will designate this
as FPD-model. For small $\eta$, where the so-called DGLAP region with $|t| >
|\eta|$ dominates, this may be justified. However, it is an open problem
how the exclusive region $|t| < |\eta|$ corresponding to the production
or absorption of a meson-like state looks like. To get a more realistic
model one can use the relation of SPDs to the so-called double distributions
(DDs) $f(z_-,z_+)$, introduced in \cite{MueRobGeyDitHor94} and rediscovered
in \cite{Rad97}
\begin{equation}
\label{DDtoSPD}
q (t, \eta, Q^2) = \int_{-1}^1 dz_+ \int_{- 1 + |z_+|}^{1 - |z_+|} dz_-
\delta ( z_+ + \eta z_- - t ) f ( z_-, z_+, Q^2 ) .
\end{equation}
The latter, according to Ref.\ \cite{MusRad99}, is given by the
product of a forward distribution $f (z)$ (more precisely $q (z)$
for quarks and $z g (z)$ for gluons) with a profile function $\pi$
\begin{equation}
\label{DD-Ansatz}
f ( z_-, z_+ ) = \pi ( z_-, z_+ ) f ( z_+ ),
\end{equation}
where $\pi$ for quarks and gluons is given by
\begin{equation}
\label{profile}
{^{Q}\!\pi} (z_-, z_+)
= \frac{3}{4}
\frac{ [1 - |z_+|]^2 - z_-^2 }{ [1 - |z_+|]^3 },
\qquad
{^{G}\!\pi} (z_-, z_+)
= \frac{15}{16}
\frac{ \left\{ [1 - |z_+|]^2 - z_-^2 \right\}^2}{[1 - |z_+|]^5},
\end{equation}
respectively. We will refer to this prescription as the DD-model. Note,
that parton distributions used here have the support $-1 \le z \le 1$
and, therefore, the contributions $- q(-z)$, $+ \Delta q (- z)$ with
$z \ge 0$ have the usual interpretation as antiquarks.

\begin{figure}[t]
\unitlength1mm
\begin{center}
\begin{picture}(150,50)(0,0)
\put(-12,-18){\inclfig{16}{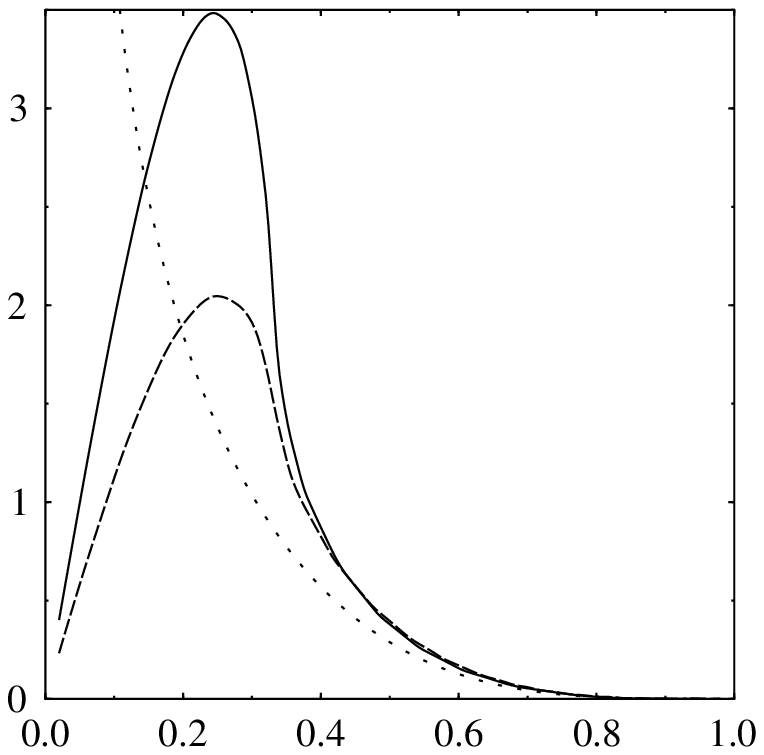}}
\put(52,45){(a)}
\put(-5,22){\rotate{${^Q\!q} (t, {\scriptstyle \frac{1}{3}})$}}
\put(36,-7){$t$}
\put(68,-18){\inclfig{16}{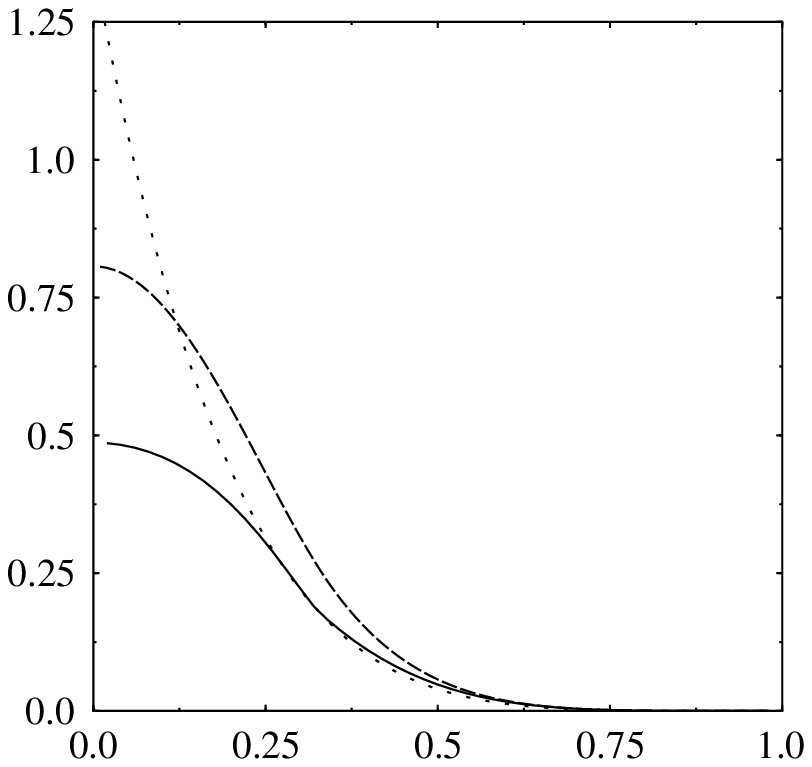}}
\put(132,45){(b)}
\put(75,22){\rotate{${^G\!q} (t, {\scriptstyle \frac{1}{3}})$}}
\put(117,-7){$t$}
\end{picture}
\end{center}
\caption{\label{FigInp}
In (a) and (b) we show the SPD for the sum of the non-polarized $u$
and $\bar u$ quarks and the gluon densities, respectively. The
FPD-model (dotted line) and the DD-model with MRS \cite{MarRobSti93}
parametrization (dashed line) are modeled at a scale of $4\ {\rm GeV}^2$,
while the DD-Model with GRV \cite{GluReyVog98} parametrization (solid
line) has been taken at the momentum scale $0.4\ {\rm GeV}^2$ and has
been evolved afterwards with NLO formulae to $\cQ^2_0 = 4\ {\rm GeV}^2$.
The skewedness parameter is $\eta = 1/3$ which corresponds to $\Bx = 1/2$.}
\end{figure}

Now we are in a position to study the NLO amplitudes in the \MS\ scheme
for the models specified above. In what follows we use the convention
$\cQ^2 \equiv - q_1^2$ and equate the factorization scale with $- q_1^2$.
The value of $\Lambda_{\overline{\rm MS}}$ is $\Lambda_{\rm NLO}^{(4)} =
246\ {\rm MeV}$. Since we will rely on available parametrizations of
the parton densities which are defined at different scales the DD-models
will differ as well. Given a forward distribution at an input scale e.g.\
$\cQ_0^2 = 4\ {\rm GeV}^2$, as e.g.\ for the MRS fit \cite{MarRobSti93},
it can be viewed as evolved only according to the DGLAP equation from a
lower scale and then folded with a profile to form the DD-model. If the
former is defined at an input of $\cQ_0^2 = 0.4\ {\rm GeV}^2$, like e.g.\
the GRV densities \cite{GluReyVog98}, to confront both models we have to
evolve GRV-based SPDs with non-forward evolution equations. We demonstrate 
these features of the DD-models in Fig.\ \ref{FigInp} where we have 
compared the MRSS0 \cite{MarRobSti93} with the GRV parametrization 
\cite{GluReyVog98} according to the procedure sketched above.

\begin{figure}[t]
\unitlength1mm
\begin{center}
\begin{picture}(150,110)(0,0)
\put(-15,42){\inclfig{16}{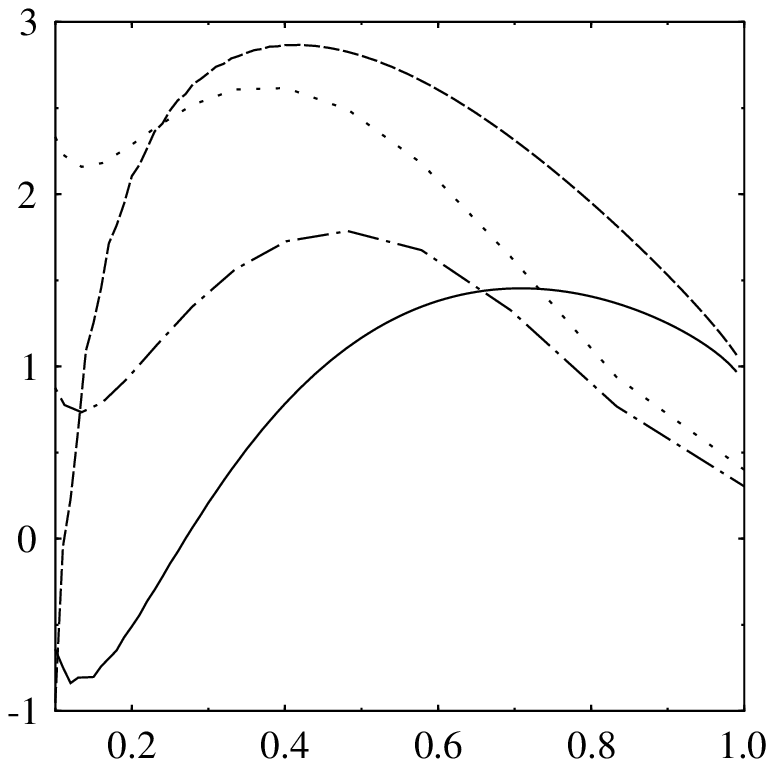}}
\put(50,67){(a)}
\put(-5,83){\rotate{${\rm Re} \cF_1$}}
\put(65,42){\inclfig{16}{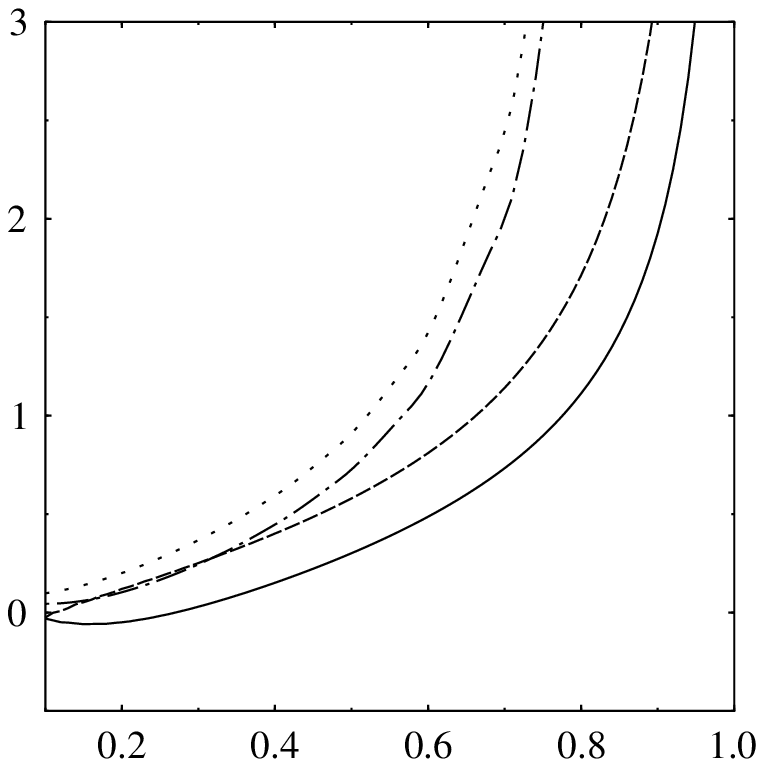}}
\put(130,67){(b)}
\put(75,77){\rotate{${\rm Re} \cF_1/{\rm Im} \cF_1$}}
\put(-15,-19){\inclfig{16}{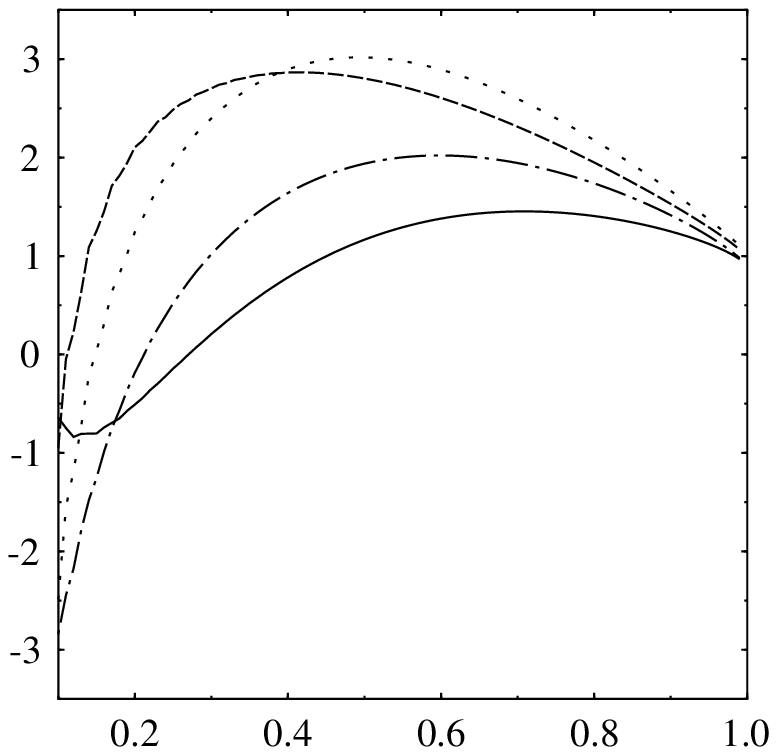}}
\put(50,6){(c)}
\put(30,-7){$\Bx$}
\put(-5,22){\rotate{${\rm Re} \cF_1$}}
\put(65,-19){\inclfig{16}{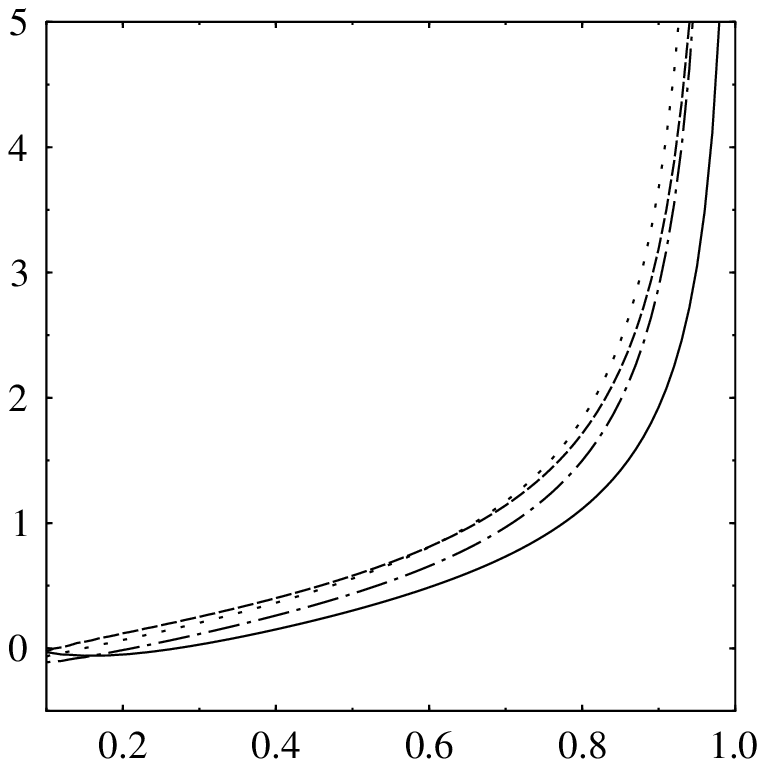}}
\put(130,6){(d)}
\put(110,-7){$\Bx$}
\put(75,17){\rotate{${\rm Re} \cF_1/{\rm Im} \cF_1$}}
\end{picture}
\end{center}
\caption{\label{pred1}
The real part (a) and the ratio of the real to imaginary part (b) of
$\cF_1$, respectively, for the DD- and FPD-model with the MRS
parametrization taken at the input scale $\cQ_0^2=4\ {\rm GeV}^2$. Here
the DD(FPD)-model at LO and NLO is shown as dashed (dotted) and solid
(dash-dotted) lines, respectively. Plots (c) and (d) show the same as
in (a) and (b), respectively but for MRS and GRV parametrizations. Here
MRS (GRV) amplitude is plotted at LO as dashed (dotted) and at NLO as
solid (dash-dotted) lines, respectively.}
\end{figure}

In general we have found that the amplitudes calculated from the models
mentioned above are similar in shape. However, in the parity even sector
the DVCS amplitude is very sensitive to the $z \to 0$ behaviour of the
sea quarks. Thus, different models will produce qualitatively different
predictions. This is demonstrated in Fig.\ \ref{pred1}(a) for $\cF_1$,
where even the sign of the amplitude for small $\Bx$ differs for the FPD-
and DD-model. Note that the sign of the FPD will change for very low
$\Bx$ and that a slightly different parametrization gives a prediction
similar to the one for a DD-model. We see from this figure that the NLO
corrections are as large as 50\% and even more. Evidently, the
imaginary part is more sensitive to the shape of the model distribution
than the real part. Their ratio is shown in Fig.\ \ref{pred1}(b). In
Fig.\ \ref{pred1}(c) we compare the predictions of the DD-model for the MRS
and the GRV parametrizations taken at different input scales as explained
above. This affects the size of both the real and imaginary part, so that
the ratio only slightly changes with the scale as shown in Fig.\
\ref{pred1}(d). The ratio is also not sensitive to the radiative corrections
in the coefficient functions. Note that the evolution will suppress
the DGLAP region, so that the asymptotic distributions are concentrated
in the region $|t| \le \eta$ and are given by the terms with the lowest
conformal spin in the conformal expansion of SPD. Thus for asymptotically
large $\cQ^2$ the imaginary part has to vanish and the radiative
corrections to the real part are determined by the lowest Wilson
coefficient in the COPE.

\begin{figure}[t]
\unitlength1mm
\begin{center}
\begin{picture}(150,110)(0,0)
\put(-15,42){\inclfig{16}{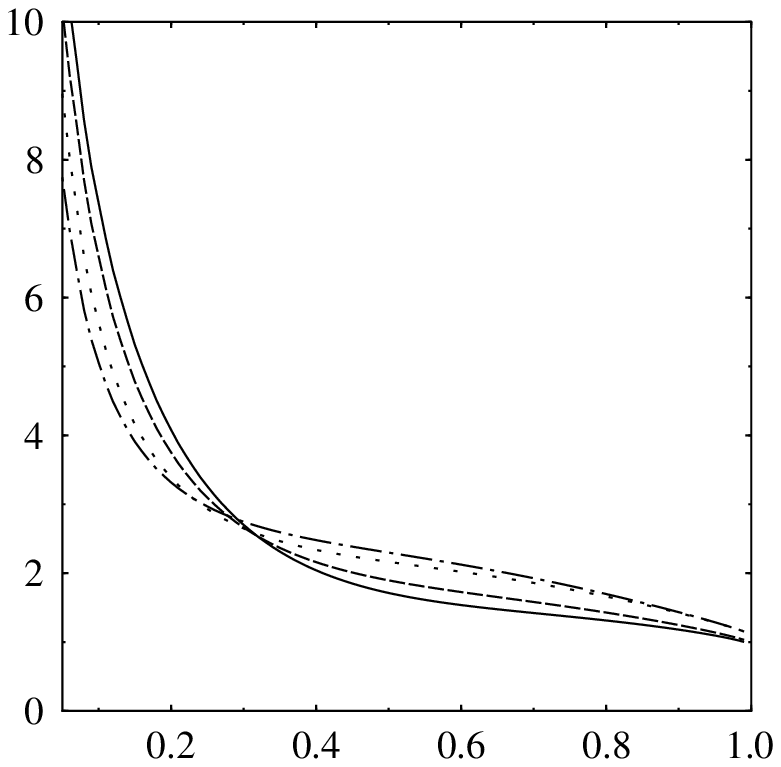}}
\put(50,105){(a)}
\put(-5,83){\rotate{${\rm Re} \cG_1$}}
\put(65,42){\inclfig{16}{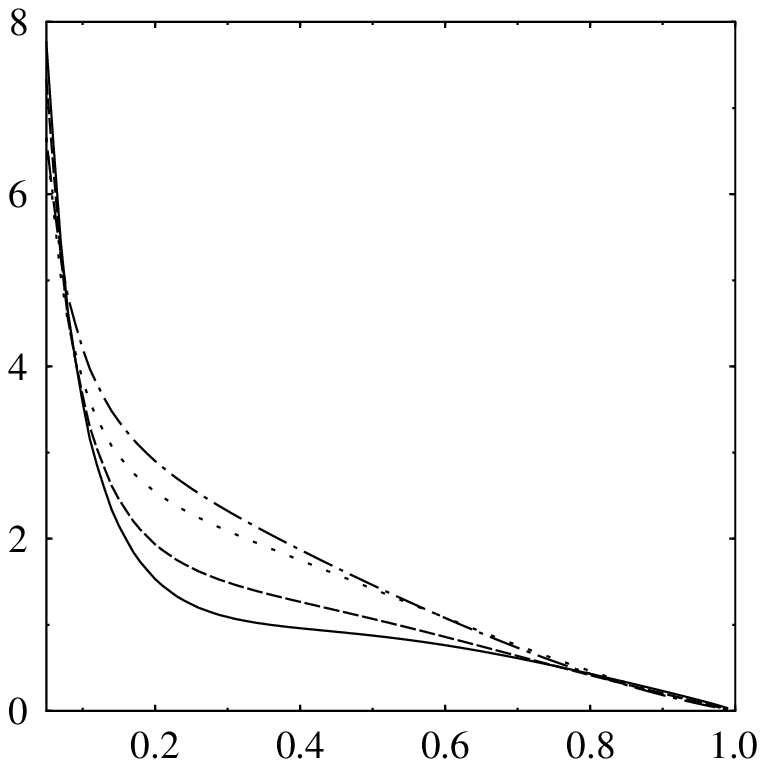}}
\put(130,105){(b)}
\put(75,83){\rotate{${\rm Im} \cG_1$}}
\put(-15,-19){\inclfig{16}{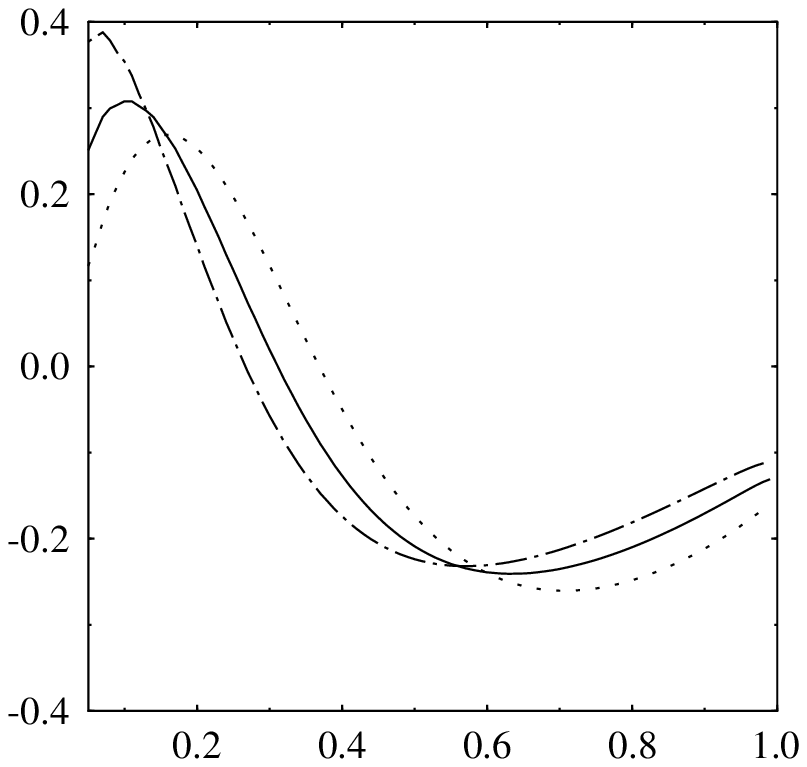}}
\put(50,44){(c)}
\put(30,-7){$\Bx$}
\put(-5,22){\rotate{$\delta {\rm Re} \cG_1$}}
\put(65,-19){\inclfig{16}{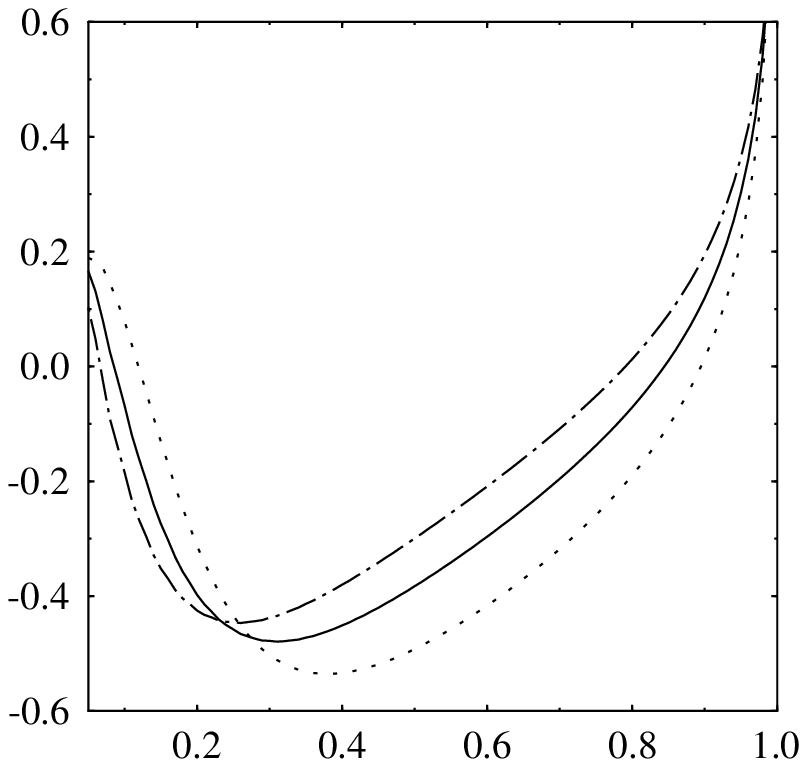}}
\put(130,44){(d)}
\put(110,-7){$\Bx$}
\put(75,22){\rotate{$\delta {\rm Im} \cG_1$}}
\end{picture}
\end{center}
\caption{\label{scheme}
The real (a) and the imaginary (b) part of $\cG_1$ for the DD-model with
the GS \cite{GehSti95} parametrization taken at the input scale $\cQ_0^2
= 4\ {\rm GeV}^2$ [LO (dotted line) and NLO (solid line)] as well as
evolved to $\cQ^2 = 10\ {\rm GeV}^2$ [LO (dash-dotted line) and NLO
(dashed line)]. Plot (c) and (d) shows the relative radiative corrections
$\delta =$ [(NLO $-$ LO)/LO] again for the real and imaginary part,
respectively, for $\cQ^2 = 4\ {\rm GeV}^2$. The factorization scale $\mu^2$
is set equal to $\cQ^2/2$ (dotted line), $\cQ^2$ (solid line), and $2
\cQ^2$ (dash-dotted line).}
\end{figure}

In Fig.\ \ref{scheme} we show the predictions for $\cG_1$ with the GSA
parametrization \cite{GehSti95} and investigate the size of radiative
corrections in detail. In this case we have found a similar model
dependence as in the previous case: predictions are sensitive to the sea
quark parametrization which turn on at $\Bx \sim 0.1$ and also may
cause very substantial radiative corrections. Since the polarized
sea quark distributions are not well known, we simply neglect them. The
real and imaginary parts for the DD-model are shown in Fig.\
\ref{scheme}(a,b), respectively, at the input scale $4\ {\rm GeV}^2$ and
evolved upwards to the scale $10\ {\rm GeV}^2$. Not surprisingly the size
of the radiative corrections is decreasing with increasing $\cQ^2$. Note
that for large $\Bx$ the ratio of real to imaginary part provides again a
useful information about the SPD as discussed above. The size of
radiative corrections for $\cF_1$ and $\cG_1$ are similar. In Fig.\
\ref{scheme}(c,d) we demonstrate the factorization scale dependence, where
we took in the hard scattering amplitude and SPD the scale to be $\mu^2 =
\{\cQ^2/2, \cQ^2, 2 \cQ^2\}$. We have also studied the perturbative
corrections in the conformal subtraction scheme, however, we have found
that in the DVCS kinematics this scheme only slightly reduces the NLO
corrections.

To summarize, we have studied the DVCS amplitude in perturbative QCD for
the region $\Bx \ge 0.05$. It turns out that a separate measurement of
real and imaginary part in this kinematic domain would provide us with
information about the SPDs, which is indispensable to discriminate between
different models. Moreover, the amplitude at $\Bx \stackrel{<}{\sim} 0.1$
depends in a crucial way on both the model and the used parametrization of
the forward parton distributions, especially, of the sea distribution.
Generally, we found that the radiative correction can be as large as
50\% and more, depending on $\Bx$. Fortunately, the ratio of real to
imaginary part is less sensitive to these radiative corrections and is
only mildly scheme dependent. Thanks to this property it can give us more
insights into the structure of SPDs. 

The kinematical situation relevant for HERA experiments requires an
extension of our analysis downwards to very low $\Bx$. In this case
the polynomial method used here will not be the most appropriate tool.
Fortunately, a direct numerical convolution of the hard scattering part
and skewed parton distributions can be carried out without major
difficulties. The remaining problem is to evolve the SPD in this
kinematics, which may be solved by numerical integration of the
evolution equation. Recently, the evolution kernels at two-loop order
have been completely constructed \cite{BelFreMue99} from the knowledge
of conformal anomalies and splitting functions and extensive use of
supersymmetric constraints. This provides in future the opportunity to
study evolution effects also at very low $\Bx$ which is a very important
task in view of ongoing experiments on diffractive meson production at HERA.

To decrease the theoretical uncertainties due to radiative corrections,
it is necessary to go beyond NLO. In a first step one can include only
the hard-scattering amplitude to two-loop order accuracy. Of course, a
direct calculation will be very cumbersome. Fortunately, however, in the
conformal limit of the theory, when the QCD Gell-Mann--Low function
formally equals zero, we can rely directly on the COPE and it seems
feasible to extend this technology to NNLO. Conformal symmetry breaking
effects can be perturbatively calculated and one piece of information is
already available \cite{GodKiv97}. From the practical point of view it
should be stressed that all high hopes connected to SPDs rely on the
assumption that the higher order terms are controllable. We want to
emphasize that our approach allows to study efficiently the structure of
perturbative corrections and might be crucial to establish the same
rigour of treatment as for the forward parton densities.

\vspace{0.5cm}

This work was supported by DFG, BMBF and the Alexander von Humboldt
Foundation (A.B.).

\end{document}